\documentstyle[psfig]{mn}

\def\propta{\hbox{${\propto}\!\!\!\!^{\sim}$}}

\title[Dwarf Nova Oscillations and Quasi-Periodic Oscillations in Cataclysmic Variables: II.]
{Dwarf Nova Oscillations and Quasi-Periodic Oscillations in Cataclysmic Variables: II. A Low 
Inertia Magnetic Accretor Model.}
\author[Brian Warner and Patrick A. Woudt]
       {Brian Warner\thanks{email: warner@physci.uct.ac.za} and 
        Patrick A. Woudt\thanks{email: pwoudt@artemisia.ast.uct.ac.za} \\
        Department of Astronomy, University of Cape Town, Private Bag,
        Rondebosch 7700, South Africa}
\date{}

\begin{document}

\maketitle

\begin{abstract}

The Dwarf Nova Oscillations observed in Cataclysmic Variable (CV) stars are
interpreted in the context of a Low Inertia Accretor model, in which
accretion on to an equatorial belt of the white dwarf primary causes the
belt to vary its angular velocity. The rapid deceleration phase is
attributed to propellering. Evidence that temporary expulsion rather than
accretion of gas occurs during this phase is obtained from the large drop in EUV
flux.

We show that the QPOs are most probably caused by a vertical thickening
of the disc, moving as a travelling wave near the inner edge of the disc.
This alternately obscures and `reflects' radiation from the central
source, and is visible even in quite low inclination systems.  A possible
excitation mechanism, caused by winding up and reconnection of magnetic
field lines, is proposed.

We apply the model, deduced largely from VW Hyi observations, to
re-interpret observations of SS Cyg, OY Car, UX UMa, V2051 Oph, V436 Cen and WZ
Sge.  In the last of these we demonstrate the existence of a 742 s period
in the light curve, arising from obscuration by the travelling wave, and
hence show that the two principal oscillations are a DNO and its
reprocessed companion.
\end{abstract}


\section{Introduction}

In an earlier paper (Woudt \& Warner 2002; hereafter Paper I) we have
presented and analysed observations of Dwarf Nova Oscillations (DNOs) and
Quasi-Periodic Oscillations (QPOs) in the dwarf nova VW Hyi, to which we 
added an overview of DNOs and QPOs in other dwarf novae and in nova-like variables.

The richness of the phenomenology, in which similarities as well as gross differences
in behaviour are discernable among the various stars, suggests that although a unifying
model may exist, there must be at least one parameter that takes different values in different
systems. In this paper we explore such a model -- the Low Inertia Magnetic Accretor (LIMA), which is
described in Section 2. 
In Section 3 we consider the suggestion that QPOs are the result of vertical 
oscillations in the accretion disc. Section 4 applies the LIMA and disc oscillation
models to interpret observations of systems other than VW Hyi, including the EUV and X-ray 
DNOs seen in SS Cyg, and in Section 5 we
conclude with a list of the principal observations and interpretative results
from the two papers, and add some general remarks.

\section{The Low Inertia Magnetic Accretor}

The model of DNOs used and developed here originates in the suggestion of
Paczynski (1978) that matter accreted during a dwarf nova outburst would generate
a rapidly rotating equatorial belt on the white dwarf primary. Any magnetic field
anchored in this belt, and at least partially controlling accretion near the primary,
would result in accretion curtains and shocks in the same manner as in
the standard intermediate polar (IP) structure (e.g.~Warner 1995a) but with
a variable frequency as the belt is spun up during the high accretion phase 
and decelerates afterwards. Any primary whose intrinsic field is too weak
(less than about $10^5$ G: see Section 2.3) to enforce solid body rotation will 
be subject to such a process,
but only those systems in which the field is, or becomes (through shearing by 
the differential rotation), strong enough to control some accretion will
show DNOs by this process. The approach, therefore, should be seen as an extension
of the IP model to lower field strengths and is compatible with the essentially 
mono-periodic nature (plus reprocessing sidebands) of the IP and DNO modulations.

This model, which results from the low inertia of the equatorial belt, has been
reviewed and extended by Warner (1995b). Since then, additional observational evidence
has accrued. HST spectroscopy obtained after a superoutburst of VW Hyi indicates
the existence of a white dwarf primary with $\log g = 8.0$, $v \sin i = 300$ km s$^{-1}$
and surface temperature 22\,500 K one day after outburst, cooling to
20\,500 K 9 days later, together with an equatorial belt maintaining a temperature
$\sim$30\,000 K and $v \sin i$ $\sim$ 3350 km s$^{-1}$ for the full ten days
after outburst (Sion et al.~1996; see also G\"ansicke \& Beuermann 1996). 
The rotation rate of the equatorial belt is close to Keplerian, and the deduced
gravitational acceleration for that component is $\log g = 6.0$, indeed implying 
dominantly centrifugal support. For an accreted belt mass of $2 \times 10^{-10}$ 
M$_{\odot}$ the temperature of the belt is in agreement with the sequence of
models computed by Sion (1995). The observations are not sufficiently precise to
exclude some deceleration of the equatorial belt in the post-outburst phase.

An equatorial accretion belt has also been deduced for U Gem (Cheng et al.~1997a;
Long et al.~1993),
decreasing in temperature and area for some weeks after outburst. Comparisons of the 
observed cooling of the white dwarf primaries after superoutbursts of WZ Sge and AL 
Com show agreement with models only if the latter include a persistent fast
rotating equatorial belt (Szkody et al.~1998).

A possible detection of the effects of magnetic channeling of
accretion flow onto the primary in VW Hyi from a trucated disc  during a
superoutburst is given by Huang et al.~(1996).  They identify an inverse P
Cygni feature which they say could be caused by ``structured gas flow onto
the white dwarf'', and also conclude that their best fitting model
``indicates that the inner edge of the disc may be detached from the white
dwarf surface''.

The existence of long-lived accretion belts impacts on predictions of the temperature
and flux from the boundary layer (BL) between accretion disc and primary. The standard 
approach (Pringle \& Savonije 1979) gives a BL effective temperature in the range
$2-5 \times 10^5$ K for high $\dot{M}$ systems, but observations of X-ray fluxes and
the ionization of CV winds constrain the BL temperature to $5-10 \times 10^4$ K (Hoare
\& Drew 1991). Accretion onto a non-rotating primary should cause the accretion 
luminosity released in the disc and in the BL to be approximately equal, but the X-ray
flux measured in VW Hyi during outburst shows that the BL luminosity is only $\sim$4 percent
of that of the disc (Mauche et al.~1991); this is consistent with the low BL temperature
deduced from CV winds. Evidently the kinetic energy in the accreting gas is not
thermalised and radiated at an equilibrium rate, as assumed in the standard model, but is
stored and lost (inwards to the white dwarf and outwards as radiation) over the spin-down
lifetime of the accretion belt. This is a consequence of the Low Inertia Magnetic Accretor
(LIMA), but the quantitative effect will differ from star to star depending on the physics
of the spin-down (e.g.~the intrinsic magnetic field of the primary, the field generated
by dynamo action in the belt, the total mass accreted during outburst).

It is useful to keep in mind the energetics of a VW Hyi outburst. Pringle et al.~(1987) 
find that $2.0 \times 10^{-3}$ erg cm$^{-2}$ is received over the range 912 -- 8000 {\AA} during 
a normal outburst, which is $1.0 \times 10^{39}$ erg for a distance of 65 pc (Warner
1987). If this is due to a mass $M(b)$ accreting on to the primary, with half the available
potential energy radiated and the other half stored in a rotating equatorial belt, then 
$G \, M(1) \, M(b) / 2\, R(1)$ = $1.0 \times 10^{39}$ erg, which gives $M(b) = 2.2 \times
10^{22}$ g for a primary of mass $M(1) = 0.6$ M$_{\odot}$ (this is compatible with the 
$\log g = 8.0$ found by Sion et al.~(1996)). If 10\% of the remaining potential energy
is lost in a wind and the rest is stored in an equatorial belt of mass $M(b)$ and angular
velocity $\Omega(b)$, then ${{1}\over{2}} \, M(b) \, R^2(1) \, \Omega^2(b) \simeq 0.9 
\times 10^{39}$ erg, which gives $\Omega(b) \simeq 0.33$ rad s$^{-1}$, or a rotation 
period for the belt of $P(b) \simeq 20$ s. For a superoutburst, for which the 
observed radiated energy is 8.4 times larger (Pringle et al.~1987), the accreted mass
should be similarly larger.

The predominantly sinusoidal character of DNOs arises from two causes.
Petterson (1980) showed that reprocessing from a disc surface in
general leads to sinusoidal pulses, independent of inclination. His
analysis of the DNO phase shifts that occur during eclipse of UX UMa implies
a considerable additional contribution from near the surface of the
primary (confirmed in the UV by HST observations: Knigge et al.~1998a).
DNOs observed in HT Cas during its outbursts (Patterson 1981) point to the
same conclusion.  The sinusoidality of this `direct' component from the primary is
understandable if the accretion zones are long arcs (perhaps 180$^{\circ}$ or more
in length) running round the equatorial belt, as expected from accretion
from the inner edge of a disc.

\subsection{Period Discontinuities}

In the LIMA the sudden changes in period of the DNOs are attributed to reconnection 
events (Warner 1995b). Field lines connecting the equatorial belt to an annulus of
the accretion disc having angular velocity $\Omega_k \ne \Omega(b)$ will be wound up.
Livio \& Pringle (1992) suggest that the twist of the field is relieved by reconnection
taking place in the magnetosphere of the primary, rather than in movement of the 
footpoints of the field in the disc or stellar interior. Lamb et al.~(1983) find that
reconnection will take place after approximately one turn has taken place between the
magnetic star and the anchored field (this corresponds to a pitch angle $\gamma \sim 1$
where the field is anchored). Thus a rotation of the primary
(or equatorial belt) relative to the disc connection annulus by about 2$\pi$ radians
will break the field lines. State-of-the-art 2D modelling, extrapolated to 3D reality,
arrives at the conclusion that ``reconnection, leading to periodic (or quasiperiodic)
evolution, would be the realistic outcome of the field twisting process'' (Uzdensky, 
Konigl \& Litwin 2001).

As pointed out earlier (Warner 1995b), there are no detectable
brightness changes at the times of the period jumps.  In VW Hyi, which has
$L \sim 0.1 L_{\odot}$ towards the end of outburst, we can set any luminosity
events as having $\Delta{L} \la 0.01 L_{\odot}$.  A
typical observed `discontinuity' is a change $\Delta{\Omega} \sim 5 \times 10^{-4}$ 
within a time $\Delta{T} \la 100$ s.  If a mass
$\Delta{M(b)}$ in the equatorial belt were changing its angular
velocity $\Omega$ by this amount, the change in rotational energy
would be $\Delta{E} \sim \Delta{M} R(1)^2 \Omega \Delta{\Omega} / \Delta{T}$,  
which gives $\Delta{M(b)} \le 4 \times 10^{14}$ g.  
It is clear, therefore, that the equatorial belt as a
whole is not varying its $\Omega(b)$ in a jerky fashion.  We suggest
that there are latitudinal variations of $\Omega(b)$, decreasing with
increasing latitude towards $\Omega(1)$ of the underlying primary.
Reconnection will then produce accretion curtains feeding gas to a range
of angular velocities, distributed around the mean value $\Omega(b)$
of the belt.  The O - C diagrams of DNOs in TY PsA (Warner, O'Donoghue \&
Wargau 1989) and in SS Cyg (Cordova et al.~1980) and our observations of VW Hyi
(Paper I) do indeed show that the
jumps of $P$ can occur in both directions relative to the underlying steady
change.

The discontinuities in $P$, and the lack of any associated luminosity
events, constitute probably the most severe constraint on any
physical model of DNOs.  An attraction of the LIMA model its its ability
to feed gas almost effortlessly onto different field lines and hence
produce small changes in $P$.

The inner radius $r_0$ of the accretion disc is determined by a combination of $\dot{M}$ and 
field strength (see below). The value of $\dot{M}$ rises and falls during a dwarf nova
outburst, decreasing and then increasing $r_0$. If $\Omega(b)$ remained at the value
determined by its connections to the innermost disc annulus, it (and hence the DNO period)
would increase and decrease smoothly as $\dot{M}$ changed. However, the field winding and
reconnection process will produce discontinuities. If we suppose that at reconnection the
field lines from the primary will reconnect to the innermost disc annulus (at $r_0$), this
being where the field is the strongest, then the change of $r_0$ and the concomitant change
in rotation period $P(b) = 2 \pi / \Omega(b)$ will result in reconnection after a time
$T_R$ found from $P(b) = {1\over{2}} {1\over{P(b)}} {{{\rm d}P(b)}\over{{\rm d}t}}
T_R^2$, or

\begin{equation}
 T_R = {\sqrt{2} P(b)\over{{\dot{P}(b)}^{1/2}}}.
\label{dnotime}
\end{equation}
 
As a quantitative example we may use the observations of TY PsA (Warner, O'Donoghue \&
Wargau 1989). The DNOs were monoperiodic and increased in period $P$ from 25.2 s
to 26.6 s over the first two nights, i.e., $\dot{P} = 1.6 \times 10^{-5}$. There were
clear discontinuities of $P$ every $\sim 10^4$ s. Equation~\ref{dnotime} gives 
$T_R \simeq 9 \times 10^3$ s, which is of the correct order of magnitude.

Furthermore, equation~\ref{dnotime} provides some qualitative explanation for the
observed change of coherence of DNOs during the course of an outburst. Near maximum 
of outburst $\dot{M}$ is most stable, so $\dot{P} \rightarrow 0$ and the time between
reconnections becomes very long. At the end of outburst, $\dot{M}$ is decreasing rapidly
and $T_R$ is reduced greatly, leading to frequent reconnections and a DNO signal
of low coherence. 

Equation~\ref{dnotime} can be rewritten as

\begin{equation}
 \Delta P \simeq 2 P^2(b) / T_R
\end{equation}

where $\Delta P$ is the sudden change in DNO period after a time $T_R$. This equation
is compatible with the limited amount of observational material on DNOs. Its statistical 
properties could afford a means of testing the LIMA model.

\subsection{The Deceleration Phase}

The rapid deceleration of the equatorial belt seen at the end of outbursts (Figure 8 of Paper I) 
has energy and angular momentum implications that limit possible mechanisms. If the 
deceleration were due to internal viscosity, coupling the belt to the primary, the
energy expenditure within or at the base of the belt
would be $\dot{E} \sim -M(b) \, R^2(1) \, \Omega(b) \, \dot{\Omega}(b) \sim
3$ L$_{\odot}$ for normal outbursts and $\sim$30 L$_{\odot}$ for superoutbursts. This
energy loss occurs at a time when the total radiant energy from VW Hyi is only
$\sim$0.1 L$_{\odot}$ and the energy radiated by the equatorial belt itself, at
$T \sim 3 \times 10^4$ K, is $\sim$0.01 L$_{\odot}$. It would appear more probable,
therefore, that the deceleration is a result of magnetic coupling to matter in the
disc. We conclude below that the belt energy is transferred to kinetic motion of gas
in the disc.

\subsubsection{Coupling of the Equatorial Belt and Disc}

We look first at angular momentum transfer between the belt and the disc.
If the magnetic field is largely generated within the equatorial belt, which itself would 
rotate nearly as a solid body through the internal magnetic viscosity, but
is weakly coupled to the primary, then a simple modification is required to the standard
IP model (e.g.~Ghosh \& Lamb 1979; see also Warner (1995a), Chapter 7, whose notation
is adopted here). The rate of change of $\Omega(b)$ is 

\begin{equation}
 \dot{\Omega}(b) = {{n (\omega_s) \, N_0}\over{M(b) \, R^2(1)}}
\end{equation}

where $N_0$ is the material torque at radius $r_0$ in the disc, i.e.~$N_0 =
\dot{M} \, [ \, G \, M(1) \, r_0\, ]^{1\over{2}}$, and $n(\omega_s)$ is the torque
function which includes both material and magnetic stresses. $\omega_s$ is the 
``fastness parameter'', given by $r_0 = \omega_s^{2\over{3}} \, r_{co}$, and $r_{co}$
is the corotation radius in the disc, given by

\begin{equation}
 r_{co}^3 = {{G \, M(1)}\over{\Omega^2(b)}}.
\end{equation}

Using $R_9 (1) = 0.73 M_1^{-{1\over{3}}}(1)$ for the mass-radius relationship of
a white dwarf in the range of $0.4 \le M_1(1) \le 0.7$, we have

\begin{equation}
 \dot{\Omega}(b) = 3.09 \times 10^{-7} \, n(\omega_s) \, \omega_s^{1\over{3}} \, \dot{M}_{16}
 \, M^{-1}_{22}(b) \, M_1^{4\over{3}}(1) \, \Omega^{-{1\over{3}}}(b)\\
\label{omegadot}
\end{equation}

Typically $n(\omega_s) \sim 0.5$, but it becomes zero when spin equilibrium is reached
at $\omega_s = \omega_c \simeq 0.975$ (Wang 1996).

First we consider any dwarf nova during outburst for which the magnetic field 
generated in the equatorial belt is strong enough to control accretion flow near the
primary. For dipole geometry, this condition ($r_0 > R(1)$) gives a field strength
$B(b) > 1.5 \times 10^4 M_1^{2\over{3}}(1) \dot{M}_{16}^{1\over{2}}$ G (Warner 1995a,
equation 7.4b), but in the vicinity of the surface of the primary other higher,
and less compressible, multipole components of the field geometry may exist. Suppose
that the equatorial belt differs in rotation period by 1 s from its equilibrium period
(at which magnetic and material torques balance). Taking $\dot{M}_{16} = 50$, 
$\Omega(b) = 0.4$ rad s$^{-1}$ and $M_{22}(b) = 2$, we have $\dot{\Omega}(b) \sim
2.5 \times 10^{-6}$ rad s$^{-2}$ from equation~\ref{omegadot}. The time scale for 
establishing equilibrium $\tau_{eq} = \Delta \Omega(b) / \dot{\Omega}(b)$ is then
$\sim$3 h. Therefore, when $\dot{M}$ is high during outburst, the rotation
period of the equatorial belt does not differ greatly from its equilibrium
value (cf.~Popham 1999).

At the end of a VW Hyi superoutburst, however, when $M_{22}(b) \sim 10$ and 
$\dot{M}_{16} \sim 1$ (as in quiescence), $\tau_{eq} \sim 30$ d and the 
rapid spin-down cannot be the result of the usual disc torques. A sufficiently
powerful torque mechanism exists, however, if the rapid rotation of the belt
is able to prevent most accretion by centrifugal action. Such a ``propeller
mechanism'' is thought to act in AE Aquarii, which has a white dwarf rotating
with period 33 s and $\dot{\Omega} = -3.9 \times 10^{-16}$ rad s$^{-2}$ (de Jager
et al.~1994), giving a spin-down time $\tau \sim 1.7 \times 10^7$ y. If a mechanism
of similar power is at work during the rapid spin-down of VW Hyi, we could expect
$\tau_{eq} \sim \tau \, M(b)/M(1) \sim 10$ h, which is similar to what is observed.
The inferred energy dissipation in AE Aqr, $I \, \Omega \, \dot{\Omega} \simeq 4$ L$_{\odot}$,
is similar to what we deduce for VW Hyi and, as for VW Hyi, is not detected as radiant energy.

In AE Aqr, modelling gives a primary magnetic field of $\sim 1 \times 10^5$ G and a mass 
transfer rate $\sim 1 \times 10^{17}$ g s$^{-1}$ (Wynn, King \& Horne 1997). Accretion
onto the primary, found from X-ray emission, is only $\sim 2 \times 10^{14}$ g s$^{-1}$
(Choi, Dotani \& Agrawal 1999), which is sufficient to produce accretion regions on the 
primary and result in modulated X-ray and optical flux (Eracleous et al.~1994). If the
VW Hyi rapid spin-down is a propeller stage, we might expect to find a similarly low
$\dot{M}$ onto the primary during that phase. This is in fact the case, as we see below
from the behaviour of the EUV flux.

It is not surprising that the system gets itself into a propellering
phase:  at the end of outburst $\dot{M}$ falls rapidly, and so $r_0$ is
increased, pushing the inner edge of the disc towards $r_{co}$. The spin
of the equatorial belt only responds slowly at first to the deceleration
torque from the disc, making it easy for $r_0$ to equal and
slightly exceed $r_{co}$. At this point $\omega_s > 1$ 
and propellering is inevitable.

\begin{figure}
\centerline{\hbox{\psfig{figure=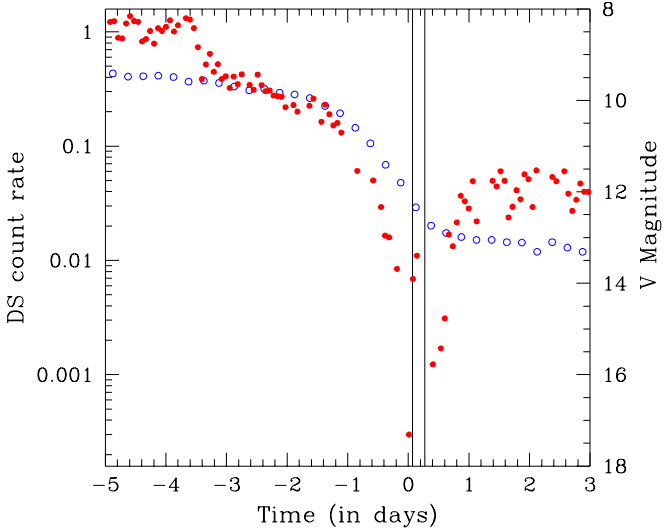,width=8.8cm}}}
  \caption{Comparison of EUV flux (filled circles) at the end of a superoutburst
with the average optical light curve (open circles). The vertical lines show the range over
which the DNOs in Figure 3 of Paper I were observed.}
 \label{euv}
\end{figure}

\subsubsection{EUV Flux as Evidence of Propellering}

It is fortunate that there are observations of the EUV flux of VW Hyi that cover
the part of the light curve in which we find the rapidly decelerating DNOs.
In Fig.~\ref{euv} we compare the EUV flux during a superoutburst of VW Hyi with the mean 
optical light curve (the phasing of the two light curves has been achieved using
AAVSO and RASNZ data obtained contemporaneously with the EUV observation). The EUV fluxes
were measured with the EUVE satellite and have been kindly communicated to us by 
Christopher Mauche (Mauche, Mattei \& Bateson 2001). Whereas the decline phase of the optical light curve
is roughly followed by the EUV flux from V $\sim 10$ to V $\sim 11$, there is a 
steepening of the EUV decline, producing a precipitous fall at V $\simeq 12.2$ and steep 
recovery starting at V $\simeq 13.0$ (this corresponds to the time of appearance of hard X-rays,
mentioned in Paper I (Wheatley et al.~1996)). This is precisely the range during which we 
see lengthening DNOs and QPOs. The EUV behaviour shows that $\dot{M}$ onto the 
primary is one or even two orders of magnitude lower at this stage than during 
quiescence.

During a normal outburst of VW Hyi the EUV flux falls even more steeply than in
the superoutburst, taking only $\sim 12$ h to fall from $\log DS$(count rate)
= -1.0 (when V $\simeq 11.0$) to $\log DS = -2.3$ (Mauche, Mattei \& Bateson 2001). The hard 
(1.5 -- 10 keV) X-ray flux at the end of a normal outburst of VW Hyi similarly 
drops to about zero when V reaches $\sim 12.5$ (Hartmann et al.~1999).

\subsubsection{Spin-down as the Result of Propellering}

The nature of the rapid spin-down, if due to propellering, can be explored in the following way. The angular
momentum drained from the equatorial belt goes into gas released at the Alv\'en radius
$R_A$. For a dipole field geometry, and gas expelled in an equatorial direction, above
and below the accretion disc,

\begin{equation}
 {2\over{3}} \dot{M}({\rm wind}) \, \Omega(b) \, R(1) \, R_A \simeq M(b) \, R^2(1) \, \dot{\Omega}(b)
\end{equation}

(Kawaler 1988), which gives, for the $\dot{\Omega}(b)/\Omega(b)$ observed in the February 2000
outburst of VW Hyi,

\begin{equation}
 {R_A\over{R(1)}} \simeq 2.5 M_{22}(b) \dot{M}_{17}^{-1}({\rm wind}).
\label{wind}
\end{equation}

It therefore requires only that most of the gas accreting from the disc be centrifuged
out along the rapidly rotating field lines to a few white dwarf radii for a large
$\dot{\Omega}(b)$ to result (this gas will not necessarily be ejected from the
system as a circumstellar wind; it will rejoin the disc at larger radii
(Spruit \& Taam 1993)). This is where the $\sim$3 L$_\odot$ extracted from the belt
is hidden: it is in kinetic energy of the centrifuged gas and is not radiated
(the same condition applies in AE Aqr).
As $\Omega(b)$ is reduced, the field $B(b)$ generated in the belt will
drop, which will reduce $R_A$ and terminate the process at some critical $\Omega_c(b)$.
In VW Hyi we have not directly observed this transition (our two observational opportunities,
on 25 Dec 1972 and 5 Feb 2000, having been terminated by natural causes), but in the
outburst of February 2000 DNOs with a period of 40.5 s were observed at the end of
our run, so $\Omega_c(b) \le 0.155$ rad s$^{-1}$ for a normal outburst. As $B(b)$ is 
probably larger for a superoutburst, making $R_A$ larger, the propeller should
be then more efficient.

It may not be necessary to think of the rapid spin-down phase as a full-blown
propeller in action. Spruit \& Taam (1993) point out that for the case where the
fastness parameter $\omega_s$ exceeds unity by only a small amount, mass piles up in the
disc and the resulting increase in pressure gradient across the magnetospheric
boundary can allow some mass flow across the centrifugal barrier (oscillation
about an equilibrium flow eventually is established, but here we are only 
considering a temporary partial damming up of the flow through the disc).
The angular momentum conveyed from the primary to this stored mass is transferred
outwards through the accretion disc.

At the end of the rapid spin-down, we see in Figure 8 of Paper I that there is either
frequency doubling (as might occur if accretion changed from a single to a two-pole
structure) or some spin-up of the equatorial belt. This phase, which occurs
at 13.0 $\la$ V $\la$ 13.3, corresponds to the rapid rise in EUV flux seen
in Fig.~\ref{euv}, and also to the second observation during a normal outburst of VW Hyi 
by the Beppo SAX satellite in the paper by Hartmann et al.~(1999), where hard
X-rays, which had decreased to about zero during the propeller phase, reappear
at $L_x \simeq 9 \times 10^{30}$ erg s$^{-1}$ and then decay to nearly a factor
of two lower in quiescence. A similar, longer term, effect is seen in the EUV
after a superoutburst (Fig.~\ref{euv}) where the flux remains higher ($\log DS$ (count
rate) $\sim$ --1.3) than the quiescent flux ($\sim$ --2.0) for some days
(Mauche, Mattei \& Bateson 2001). This latter phase corresponds to the brighter
V magnitude after superoutburst than after normal outburst (Figure 5 of Paper I).

We suggest that such a spin-up could be expected to follow the propeller phase, because
mass that was stored in the inner disc, is eventually allowed to continue accreting on to the primary. 
This phase accounts for the period of enhanced $\dot{M}$ and should
last longer after a superoutburst because the propellering was more powerful
and expelled more gas.

By the end of these spin-down and spin-up phases, the effect of the belt's magnetic field on the
inner disc will drain the latter in a short time (Livio \& Pringle 1992).
Rapid rotation of the belt may then continue at an angular velocity 
sufficient to prevent accretion in the equatorial plane. This can be seen
from the estimate that $\dot{M}(1)$ onto the primary during quiescent is 
found from the hard X-rays to be $\sim 2.5 \times 10^{14}$ g s$^{-1}$
(van der Woerd \& Heise 1987; Hartmann et al.~1999), which is in accord (a) with the
requirement that $\dot{M}(1) \ll | \dot{M}(2) |$ in quiescence of a dwarf nova
(which stores mass in its disc between outbursts), and (b) with the estimate that the mass
transfer rate $\dot{M}(2)$ from the secondary should be about twice that observed
in Z Cha (which has outbursts at half the frequency of those in VW Hyi) which
is $2 \times 10^{15}$ g s$^{-1}$ (Wood et al.~1986). 

The radius $r_{\mu}$ of the
magnetosphere in an intermediate polar (see, e.g., Warner 1995a) is 

\begin{eqnarray}
 r_{\mu} & \simeq & 1.38 \times 10^{10} \mu_{32}^{4\over{7}} \dot{M}_{14}^{-{2\over{7}}} 
 M_{1}^{-{1\over{7}}}(1) \,\,\,\, {\rm cm} \\
\nonumber & = & 13 \mu_{32}^{4\over{7}} R(1)  \hspace{0.5cm} {\rm for} \, \dot{M}_{14} = 
 2.5 \,\, {\rm and} \,\, M_{1}(1) = 0.60
\end{eqnarray}

where $\mu$ is the magnetic moment, and the corotation radius is

\begin{equation}
 r_{co} = 1.50 \times 10^{8} \, P^{2\over{3}} \, M_1^{1\over{3}}(1) \,\,\,\, {\rm cm}
\end{equation}

and therefore $r_{\mu} > r_{co}$ unless $\mu_{32} \le 2.8 \times 10^{-2}$, or
$B(1) \le 4.5 \times 10^3$ G (this may be compared with the conditions at the 
peak of outburst, where $\dot{M}(1) \sim 5 \times 10^{17}$ g s$^{-1}$ and
the requirement that $r_{\mu} > R(1)$ in order to produce the magnetically 
channelled accretion implied by the 14 s modulation gives $B(1) > 7.7 
\times 10^4$ G). Therefore, it would need an extremely weak field to allow
the magnetosphere in quiescence to be compressed to the point where accretion as an
IP is allowed. The small amount of accretion that manages to reach the
primary is evidently not predominantly magnetically channelled
(from the absence of DNOs in the optical or in the hard
X-ray flux: Belloni et al.~(1991) -- these authors only examined the region
13--15 s, looking for X-ray modulations similar to those seen at maximum
of outburst. A more catholic search in the various hard X-ray light curves, for 
periods in the range of 20--100 s, should be made.), 
and may be a result of more isotropic accretion from a hot corona,
as in the models of van der Woerd \& Heise (1987) and Narayan \& Popham (1993), 
which extends to a distance $\sim R(1)$ above the surface of the primary when the
gas becomes optically thin in the X-ray region.

\subsubsection{The Post-outburst Equatorial Belt}

The detection in VW Hyi (Sion et al.~1996) of rapidly rotating equatorial belts one day
after a normal outburst (and covering $\sim$11 percent of the surface area
of the primary) and 10 days after a superoutburst (covering reduced to $\sim$3 percent)
shows that although the belt may persist throughout the average $\sim$28~d of 
quiescence, its spin-down time scale is comparable to the outburst interval. 
Therefore the torques acting during outbursts are largely on the gas accreted
during the outburst, with little allowance required for gas left from the 
previous outburst.

The hot belt left at the end of outburst contains the thermal energy of compression
and kinetic energy of rotation which will partly be thermalised internally by viscous
stresses and partly lost by magnetic coupling to the disc. Even if there is no magnetically 
controlled accretion during quiescence, coupling still exists through field lines
connecting the belt to the slower rotating parts of the accretion disc, and through
possible continued weak propellering. The thermal energy in the belt is partly 
radiated (as seen by its temperature excess over the $\sim$18\,000 K quiescent temperature
of the primary) and partly conveyed into the primary. The latter is detected through the 
excess temperatures of the primary measured by Sion et al.

The observed $v \sin i$ for the primary of VW Hyi is 300 km s$^{-1}$ (Sion et al.~1996) which,
with $i \sim 60^{\circ}$, gives a rotation period of the primary of $P(1) \simeq 157$ s.
The kinetic energy in the belt at the end of the propeller phase is 
${1\over{2}} M(b) R^2(1) \{ \Omega(b) - \Omega(1) \}^2$, which is 
$\sim 5 \times 10^{38}$ erg if $P(b) \sim 40$ s and $M_{22}(b) \sim 5$.
All of this must be lost in $\sim$25 d (in order to maintain a steady state), 
requiring an average luminosity of 0.06 L$_{\odot}$, or 4.5 times the luminosity
of the steady state 18\,000 K primary (Long et al.~1996). This energy loss
is readily accounted for by the
observed cooling of the belt and primary. Some of the energy will of course be
transferred to the interior through viscous coupling of the belt to the interior. 
In VW Hyi this coupling is evidently quite weak, possibly implying a relatively 
weak magnetic field in the primary -- which is slightly ironic in view of the fact
that VW Hyi supplies so many observable magnetically related phenomena at the
end of its outbursts.

\subsection{Magnetic Modelling}

Here we will give a few plausibility arguments for the LIMA model.

\subsubsection{Transmission of Accretion Stress to the Interior}

To couple the exterior of a star to its interior requires a process that
can transfer stresses. A non-magnetic, non-crystalline core of a white dwarf
requires $\ga 5 \times 10^{10}$ y to transport angular momentum (Durisen 1973), but
the high stability of the rotation of DQ Her shows that magnetic accretors are
able to couple surface torque to the deep interior. As pointed out by 
Katz (1975), following the early work of Mestel (1953), a relatively small magnetic 
field is able to provide the means of coupling. This can be seen as follows.

The stress $R {\rm d}(\rho v)/{\rm d}t$ of accretion can be transported to the
interior by the Maxwell stress of field lines, if $B^2/4 \pi$ is considerably 
larger than the mechanical stress. That is,

\begin{equation}
B^2 \gg 4 \pi \rho \dot{\Omega} R^2(1)
\end{equation}

or
\begin{equation}
B \gg 14 \rho^{1\over{2}} \dot{M}_{17}^{1\over{2}} M_1^{-{5\over{6}}}(1) \,\, {\rm G}
\end{equation}

where we have made similar approximations to those used in earlier
Sections.

Quite small fields are therefore capable of stiffening the outer regions
of white dwarfs (in particular, causing almost solid body rotation in an
equatorial belt), but for $\rho \simeq 10^6$ g cm$^{-3}$, which in an intermediate
mass white dwarf occurs at a depth where $\sim$20\% of the mass has been
traversed, it requires $B \gg 2 \times 10^4$ G to couple to the rest of the
interior. Fields $\ga 10^5$ G therefore are required to transport angular 
momentum throughout the star\footnote{Note that the exceptional spin-down
power of AE Aqr (de Jager et al.~1994) is based on the assumption that the
primary rotates as a solid body. The field strength $B \sim 1 \times 10^5$ G deduced
by Choi and Yi (2000) would suggest that this is not necessarily the case: it is
possible that only the outer parts are being spun down.}.

\subsubsection{Field Enhancement by the Equatorial Belt}

An estimate of the enhancement of the intrinsic field of the primary
caused by accretion around the equator may be obtained as follows.
The radial component $B_r(1)$ of the primary is converted into 
toroidal field $B_{\phi}(b)$ of the belt as the field lines are carried around by 
the conducting fluid flow. In one rotation of the belt, a field strength 
$B_{\phi}(b) \simeq B_r(1)$ is generated (e.g., Livio \& Pringle 1992)
and this increases proportional to the accumulative angle (we assume
here that $\Omega(b) \gg \Omega(1)$). Magnetic flux is lost on a time
scale $\tau$, and hence the mean field strength is

\begin{equation}
B_{\phi}(b) \simeq B_r(1) {{\tau}\over{P(b)}}.
\end{equation}

If the loss mechanism is bouyancy of the field, rising at the Alfv\'en velocity $v_A$, then
$\tau \simeq H / v_A$, where $H$ is the scale height and 
$v_A = B_{\phi}(b) \, (4 \pi \rho)^{-{1\over{2}}}$, which gives

\begin{equation}
B_{\phi}^2 (b) \simeq {{\Omega(b) \, H}\over{\pi^{1\over{2}}}} \, \rho^{1\over{2}} \, B_r (1).
\end{equation}

Using $\rho H = M(b) / 2 \pi x R^2(1)$, where $x R(1)$ is the (latitudinal) width of the
accretion belt, and an effective gravity of $g_{eff} = y g$, where $g = G M(1) / R^2(1)$, we find

\begin{equation}
B_{\phi} (b) \simeq 210 B_r^{1\over{2}}(1) P^{-{1\over{2}}}(b) \Big( {{M_{22}(b) T_5}\over{x y M_1(1)}} \Big)^{1\over{4}} \,\,\,\, {\rm G}
\label{bphi}
\end{equation}

where $B_{\phi} (b)$ and $B_r$ are measured in Gauss, $P(b)$ in secs and $T_5$ is the temperature in
the belt in units of $10^5$ K.

Using $x=0.1$, $y=0.01$ (as indicated by $\log g = 6.0$, found by Sion et al.~1996), we
have, for $P=20$ s, $M_{22}(b) = 4$, the field strengths as given in Table~\ref{tab2},
and therefore some field enhancement appears feasible.  The geometry of a rotation-enhanced
field is probably quite complicated. The $B_{\phi}$ component estimated here will determine
the radius of the inner edge of the accretion disc, and any region with a strong radial
component will allow accretion onto the primary. This same radial field is responsible for
propellering.

\begin{table}
\caption{Field strengths in the equatorial belt}
\begin{tabular}{lrrr}
$B_r(1)$        & 10$^3$ G & 10$^4$ G & 10$^5$ G \\
$B_{\phi}(b)$   & $1.2 \times 10^4$ G & $3.8 \times 10^4$ G & $1.2 \times 10^5$ G \\
\end{tabular}
\label{tab2}
\end{table}

\subsubsection{Predicted and Observed $P$-$\dot{M}$ Relationships}

Magnetically channelled accretion flow at the inner edge of the accretion disc
requires that the magnetospheric radius of the primary be larger than the stellar radius.
This condition may be written (e.g.~equation 7.4b of Warner 1995a)

\begin{equation}
B(b) \ga 4.7 \times 10^4 \, M_1^{2\over{3}}(1) \, \dot{M}_{17}^{1\over{2}} \,\,\,\, {\rm G}
\hfill {\rm for \, a \, dipole \, field}\,\,\,\ 
\label{di}
\end{equation}

\begin{equation}
B(b) \ga 9.0 \times 10^3 \, M_1^{2\over{3}}(1) \, \dot{M}_{17}^{1\over{2}} \,\,\,\, {\rm G}
\hfill {\rm for \, a \, quadrupole \, field}\,\,\,\ 
\label{qua}
\end{equation}

For accretion from as close to the primary as in dwarf novae in outburst, any quadrupole
component of the field geometry is therefore likely to dominate (see also Lamb 1988). Comparison
of Table 1 and equations~\ref{di} and \ref{qua}, indicates that magnetically controlled accretion into
the equatorial belt is plausible even for intrinsic fields $B_r(1)$ as small as $10^3$ G.

In the LIMA model the predicted DNO periods are

\begin{equation}
P \propto \mu^{6\over{7}} \, \dot{M}^{-{3\over{7}}} \hfill {\rm dipole}\,\,\,\ 
\label{dip}
\end{equation}

\begin{equation}
P \propto \mu^{6\over{11}} \, \dot{M}^{-{3\over{11}}} \hfill {\rm quadrupole}\,\,\,\ 
\label{quap}
\end{equation}

(Warner 1995b). Mauche (1997) has shown that in SS Cyg, the DNO periods behave
more as $P \propto \dot{M}^{-{1\over{10}}}$ (where $\dot{M}$ is obtained from the
EUV flux). This weak dependence on $\dot{M}$ is readily understandable with the
LIMA model.

First we note that, with $M(b)$ held constant, equations~\ref{bphi}, \ref{dip}, \ref{quap}
predict (since $\mu \propto B \propto P^{-{1\over{2}}}$ in equation~\ref{bphi})
$P \propto \dot{M}^{-{3\over{10}}}$ for dipole and $P \propto \dot{M}^{-{3\over{14}}}$ for
quadrupole. But $M(b)$ increases during an outburst, as $M(b) = \int_0^t \dot{M} {\rm d}t$, 
and this in equation~\ref{bphi} enhances $B_{\phi}(b)$ which acts to decrease
further the dependence of $P$ on $\dot{M}$. It is, for example, possible to 
obtain $P \, \propta \, \dot{M}^{-{1\over{10}}}$ by allowing $\log \dot{M}$ to vary approximately
parabolically with time, as in the models of SS Cyg computed by
Cannizzo (1993).

\section{The Quasi-Periodic Oscillations}

In this section we explore the suggestion that the QPOs are the result of vertical
oscillations of selected annuli of the accretion disc.  As already
outlined in Section 3 of Paper I, the reason for considering such oscillations
is that they can provide luminosity modulations at periods much longer
than those of the DNOs.

\subsection{Location of the QPOs}

The systematic study of wave propagation in accretion discs given by
Lubow \& Pringle (1993; hereafter LP) and the numerical results given by
Carroll et al (1985) show that QPOs with periods greater than about 100 s 
arise from nonaxisymmetric modes.  In particular, it is the $m = 1$
prograde (i.e., moving in the same direction as Keplerian motion, as
viewed from an inertial frame) modes which can be excited.  These modes
are equivalent to a vertical thickening of the disc over an extended range
of azimuth which travels retrogradely in the frame of rotation of the
disc, but at a slightly lower angular frequency than that of the Keplerian
rotation.  Figure 12c of LP shows that low frequency retrograde waves are
unable to propagate (in LP's notation, $K^2_x < 0$ for
small $F$) but low frequency prograde modes waves can propagate ($K^2_x > 0$ for 
small $F$ in Figure 12b).  LP's final conclusion is
that ``the $m = 1$ g-mode appears to provide the best propagational
properties near the disc centre''.

To generate oscillations an excitation mechanism is required, which
must be selective in order to produce the distinctive time scales seen in
individual CVs.  There are three radii within an accretion disc in which
excitation mechanisms might arise:  (a) the annulus in the disc at which
any stream overflow impacts onto the disc, (b) the inner edge of the disc,
near radius $r_0$, and (c) the outer edge of the disc where stream impact
occurs.

The stream overflow impact on the inner disc occurs close to the
stream's radius $r_{min}$ of closest approach to the primary (Lubow \& Shu 
1975; Lubow 1989) which, as a fraction of the separation of the
centres of mass of the stellar components, is at

\begin{equation}
 {r_{min}\over{a}}  =  0.0488 q^{-0.464}
\end{equation}

(Warner 1995a) where $q$ is the mass ratio (secondary mass $M(2)$ divided
by primary mass $M(1)$).  The Keplerian velocity is

\begin{equation}
 v_k(r)  = \big( {{G M(1)}\over{r}} \big)^{1\over{2}}
\end{equation}

from which, with Kepler's Third Law, we obtain the Keplerian period as a
fraction of the orbital period:

\begin{equation}
P_k (r_{min}) / P_{orb}  =  0.0107 q^{-0.70} (1 + q)^{1\over{2}}.
\end{equation}

For a star such as VW Hyi, with $q \simeq 0.15$ and
$P_{orb} \sim 100$ min, we have $P_k (r_{min}) \sim 300$ s.  It is
therefore conceivable that QPOs with periods of hundreds of seconds could
be excited by the stream overflow process.

Our observations of VW Hyi in February 2000, however, show that there is an
evolution of QPO period which parallels that of the DNOs, and these are
different from those observed in December 1972.  QPOs excited by stream impact 
would be expected to have a fixed time scale.  Furthermore, stream impact
(which occurs at a greater velocity than the local Keplerian velocity) is
unlikely to excite waves that are retrograde in the Keplerian rotating
frame (as appears to be required by some of our observations).

Although some of the QPO periods in CVs are comparable to the Keplerian
period at the outer edge of the disc, and might be oscillations excited by the initial
stream impact (at the bright spot), the periods of 250 -- 500 s seen in VW
Hyi are too short ($P_k / P_{orb} \sim 0.2$ at the outer edge of the
disc (equation 8.6 of Warner 1995a) which is $\sim$ 1300 s for VW Hyi), and the
evolution towards longer periods seen at the end of outburst occurs at a
phase when the outer radius of the disc is known to be shrinking (e.g., in
Z Cha:  O'Donoghue 1986), which would produce decreasing periods.

We therefore turn to consideration of oscillations located near the
inner edge of a truncated disc.

\subsection{Inner Disc QPOs and their interaction with DNOs}

For the low field white dwarfs in dwarf novae the inner edge of the disc
is close to the primary, so opportunities for a bulge to intercept
radiation from the central source are relatively high.  In particular, a
travelling wave running around the inner disc will intercept some of the
constant luminosity component of the primary, producing (a) modulation as
its illuminated surface is seen at different aspects, (b) modulation as it
obscures radiation from the central source (as seen by us), and (c) a
shadow cast over the illuminated concave surface of the accretion disc.
Interruption and obscuration will also affect the rotating beam which is the source of
the DNOs.  The relative importance of these effects will depend partly on orbital
inclination - e.g., they will have a maximum effect at high inclination.
At low inclination all but one of the effects should disappear:  there is
still the possibility of a travelling bulge intercepting the rotating beam
and generating optical DNOs (which will be at a lower frequency than any
simultaneous X-ray DNOs, see below). The contributions of these several
components are likely to vary rapidly with time, as the amplitude and
profile of the travelling bulge changes. This accounts for the wide range 
of behaviour in the interactions between the DNOs and QPOs described in Section
4.2.3 (and Figures 20 and 21) of Paper I.

\subsubsection{Production of QPOs by a Travelling Wave}

We consider the interception of radiation from the primary by a vertical
thickening of the disc caused by a travelling wave.
To produce an amplitude $\Delta{m}$ in the light curve requires a wave
of height $h \sim \pi \Delta{m} r_0$ if the wave has a sinusoidal profile, 
radiation is emitted uniformly over the surface of the primary, reprocessing
is highly efficient, and we see all of the wall. This last requirement --
that the primary does not obscure any of the far side of the wall -- translates
to $r_0/R(1) \ge (1 - (h/{r_0}) \tan i)^{-1} \sec i$. Taking $i \simeq 60^\circ$
for VW Hyi we have $r_0/R(1) \ge 2.1$ for $h/r_0 = 0.10$. As we are mostly dealing
with QPOs observed during the propellering phase, where the inner edge of the disc
is at a few times $R(1)$ (see equation \ref{wind}) this condition is satisfied
for even quite high QPO walls. For higher inclination systems, or for smaller $r_0$
in VW Hyi, reprocessing from the far side of the wall may be largely obscured by 
the primary.

The mean amplitude of QPOs seen in Figure 1 of Paper I is $\simeq 0.05$ mag which, 
from the above, requires $h/r_0 \simeq 0.15$. At this time the primary in VW Hyi
has a temperature of at least 25\,000 K (G\"ansicke \& Beuermann 1996), i.e.,
$L \ge 2.0 \times 10^{32}$ erg s$^{-1}$. The fraction of this, if radiated isotropically,
intercepted by the wall is $\sim h/\pi r_0 \sim 1.0 \times 10^{31}$ erg s$^{-1}$.
At the end of outburst, VW Hyi has $L \sim 4 \times 10^{32}$ erg s$^{-1}$ in the
optical region (Pringle et al.~1987). If reprocessing into the optical is very efficient
we would therefore expect an amplitude $\sim 0.025$ mag. However, in reality most of the
radiation comes from the hot equatorial belt, which is both closer to the wall and more 
luminous than most of the surface of the primary, so QPO amplitudes of 0.05 mag or more are
energetically feasible.

The roughly constant amplitude (on a magnitude scale) of the QPOs in 
Figure 1 of Paper I shows that a constant fraction of the radiation emitted
by the central source is being periodically intercepted. It suggests
that at this late phase of the VW Hyi outburst the brightness in the optical
is largely due to reprocessed radiation from the hot primary, rather than 
intrinsic luminosity of the accretion disc. This is also seen by the very large
amplitude of the QPOs in the 12 Sep 1972 observation (Figure 12 of Paper I), where reprocessing
from the travelling wall at times accounts for a large fraction of the system 
brightness. Note also that at the end of a superoutburst the fall to quiescent magnitude
is slower than in normal outbursts (Figure 5 of Paper I). This may be due to the larger
decay time of irradiation by the more massive equatorial belt.

\begin{figure}
\centerline{\hbox{\psfig{figure=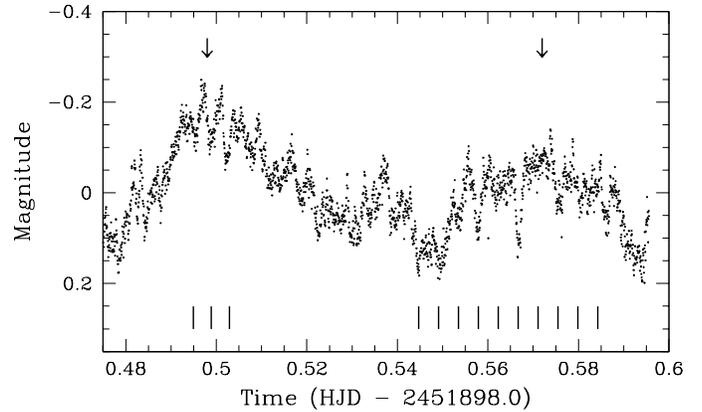,width=8.8cm}}}
  \caption{Part of the light curve of VW Hyi obtained on 19 December 2000. Prominent periodic dips are marked by 
vertical bars. Arrows indicate predicted times of orbital hump maxima.}
 \label{dips}
\end{figure}

Alternate reprocessing and obscuration is particularly marked in the 
section of the 19 Dec 2000 run shown in Fig.~\ref{dips}. The minima of the QPO modulation
clearly pull the system brightness well down below the smoothed average, both
on and between the orbital humps. Notice in particular in the upper panel of Figure 12 of Paper I 
that a large dip between
the orbital humps is immediately followed by a large peak -- evidently the wall was of
exceptional height for that revolution; the central part of the light curve on 6 November 1990
(Figure 12 of Paper I) also shows this, where the obscuring wall survives for several revolutions. Two short lived
periodic dips can also be seen in Fig.~\ref{dips}; one around HJD 2451898.5 at a period 342 s for about 3 cycles,
and another QPO modulation in the last part of the run at 382 s. In this latter QPO, deep dips are seen
at twice the QPO period, i.e.~at 760 s, cutting well into the orbital hump (marked by the arrows).

Periodic obscuration of the central source by a travelling bulge
should lead to a periodic component in the absorption spectrum.  It is
possible that in VW Hyi this has already been seen: Huang et al (1996)
observed VW Hyi at supermaximum with the HST and found deep narrow
absorption cores in N V and C IV lines in one spectrum, which had  
disappeared in a second spectrum taken ten minutes later.  A prediction of
our model is that such variation in the column density, if caused by the
rotating bulge in the inner disc and not by, e.g., absorption by the
accretion stream, should be modulated at the QPO period.

\subsubsection{Interception of the DNO Beam by the Travelling Wave}

Reprocessing of the rotating beam by the travelling bulge generates a 
second set of DNOs whose frequency is the difference between the beam
and bulge frequencies. That is, at a reprocessed period $P_{DNO'} =
(1/P_{DNO} - 1/P_{QPO})^{-1}$. This is the relationship we have found in
VW Hyi (Figure 18 of Paper I). The waveform of the reprocessed DNOs will be a 
convolution of the profile of the DNO beam and that of the bulge. 
If the bulge is non-sinusoidal then the reprocessed profile will have
harmonics in it. This is seen in our observations. There is, however,
another possibility. Suppose that there are two accretion zones, as in
dipolar accretion, with one above the equator and the other (perhaps 180
degrees away in longitude) below the equator.  The rotating beam from the
former will illuminate the upper surface of the disc, generating optical DNOs
observable by us, but the latter illuminates the under surface of the disc
which is invisible to us.  The travelling bulge, however, may intercept
radiation from both accretion zones, reprocessing it to become a modulated
optical signal visible to us (when the bulge is at `superior
conjunction').  This signal will in general show both a fundamental and
first harmonic, arising from the different intercepted fluxes of the two rotating
sources of illumination.

\subsection{Excitation of QPOs}

Although eigenmode analyses of complete accretion discs have been made
(see Section 3 of Paper I), excitation mechanisms in magnetospherically truncated
discs do not appear to have been explored.  Here we suggest a possible
mechanism, based on the standard IP model, which may excite modes of the
correct period and which also seems capable of explaining why the ratio of
DNO to QPO period remains roughly constant during our observations of VW
Hyi made in February 2000.

In the model of accretion from a disc into a magnetosphere (Ghosh \&
Lamb 1979) outlined in Section 2, there are three radii of importance: the
largest is the corotation radius $r_{co}$ at which the angular velocity
$\Omega(1)$ of the magnetosphere matches the Keplerian velocity; then
comes $r_0$ at which the magnetosphere begins to influence the fluid
flow in the disc, where for $\omega_c$ = 0.975,  $r_0 =
0.975^{2\over{3}} r_{co} = 0.983 r_{co}$ for the equilibrium
state; finally there is $r_A$ where the disc flow matches the
magnetospheric rotational velocity and accretion along field lines becomes 
possible.  The region contained within $r_A \le r \le r_{co}$ is therefore 
one where the disc flow
speed is greater than the speed of the field lines.  The field lines
consequently are dragged forward by the plasma, producing a spin-up torque
on the primary or its equatorial belt.  Conversely, we can think of that
region (Ghosh and Lamb's `boundary layer') as one where the torque from
the magnetosphere decelerates the accretion flow until it matches the
magnetospheric angular velocity.  This deceleration zone (DZ) is a
plausible site for exciting disc oscillations.

Consider equilibrium accretion (i.e. with $\dot{M}$ constant).  In general
the primary or its equatorial belt will have an inclined magnetic axis (an
aligned, symmetrical field would produce annular accretion zones around
each rotation pole and not generate the anisotropic radiation pattern
required for observable DNOs).  There will be an azimuthal variation of
field strength around the DZ. The strongest field lines threading the DZ
will occur at a particular azimuth. As the plasma attached to those field
lines advances relative to the rotation flows at $r_A$ and $r_{co}$ on
either side of it, it drags the field lines forward, is decelerated in the
process and piles up and is shocked by gas following in the same annulus
threaded by weaker field lines.  This heats and thickens the disc in this
region.  Gas may also be lifted out of the disc in this region as the
field lines are bent over by the twisting motion and provide some vertical
magnetic pressure support.  The result is a region of higher density 
and greater vertical thickness (a `blob', to adopt a
frequently used technical term).

The growth of the blob terminates when the field lines threading it have
advanced so far relative to the $\Omega(1)$ rotation that they break
and reconnect to the disc in an energetically more favourable position.
The sudden release of the blob from its decelerated and elevated state
will provide the perturbation required to launch a travelling wave at a
speed determined by the local conditions in the annulus. This wave may
damp out quite quickly, but it will be regenerated at the next
reconnection event.  The time scale of wave generation will then be
determined by the recurrence period of reconnections;  there is
possibility of a resonance which would generate waves of considerable amplitude.

The time scale of reconnection can be estimated as follows.  Let the
angular velocity of the plasma in the region of strongest field threading
be $\Omega(max)$.  If reconnection occurs when the field lines have
advanced through an angle $\theta$ relative to their untwisted
position, then the time $T_R$ between reconnections is given by

\begin{equation}
T_R [\Omega(max) - \Omega(1)] = {\theta}.
\end{equation}

Typically $\theta \sim 2\pi$ (Lamb et al.~1983), and Ghosh \& Lamb
(1978) show that the angular frequency $\Omega(r)$ is maximum at $r =
r_0$, therefore $\Omega(1) < \Omega(max) < \Omega(r_0)$.  
The value of $\Omega(max)$ appropriate for the present need is not available 
from existing models of the DZ:  in the models given by Ghosh \& Lamb,  
the greatest magnetic dissipation and consequent deleration occurs close to the inner boundary
ar $r_A$, but the details of this model are disputed by Wang (1987).
  
If we identify $T_R$ as the time scale that will resonate with the
disc oscillation, and adopt $\Omega(max) \simeq \Omega(r_0)$, we have

\begin{equation}
{P({\rm QPO})\over{P({\rm DNO})}}  =  {\omega_s \over {1 - \omega_s}}.  
\label{qpodno}
\end{equation}

The observed ratios $P$(QPO)/$P$(DNO) are in the range 10 - 20, implying
values of $\omega_s$ in the range 0.91 - 0.95.

Equation~\ref{qpodno} has the property that, for $\omega_s \simeq$ constant,
$P$(QPO) \propta \, $P$(DNO), as is seen in VW Hyi in the
February 2000 observations.
  
It may be noted that the winding and reconnection of field lines
described here should also occur in the intermediate polars.  With typically
$P$(1) = 15 minutes, equation~\ref{qpodno} shows that any resonant QPO would have a
period of $\sim$ few hours, which is comparable to the orbital period and
would be difficult to detect photometrically.

In systems of low orbital inclination, luminosity modulation at the revolution
period of the travelling wall would not be expected -- but the growth and decay of
a wall, reprocessing radiation into our direction, could lead to QPOs with mean
periods equal to the average {\it lifetime} of the wall.

In some of our light curves, particularly Figure 1 of Paper I, we have the impression
that, apart from photon counting and scintillation noise, almost all of the observed
rapid variations are due to DNO and QPO modulation -- the rapid flickering associated
with CVs is weak or absent. If the normal flickering activity is associated with
magnetic instabilities in the inner disc (e.g.~Bruch 1992), this may indicate that the 
presence of QPOs, requiring a more organised magnetic structure, supresses
the flaring activity.

Unless there is another, nonmagnetic, excitation mechanism for QPOs, our 
observations of occasional QPOs during quiescence (Section 4.2.2 of Paper I) shows 
that some magnetic disc--primary interaction can occur at minimum light, even
if little or no magnetically controlled accretion occurs.

\section{Application to other systems}

Several of the DNO and QPO phenomena that we observe in VW Hyi are present in
other cataclysmic variables. The models that we have developed for VW Hyi throw some
light on what is happening in these other systems.

\subsection{X-ray and EUV DNOs in SS Cyg}

The only DNOs in VW Hyi observed in the X-ray region are those analysed by van der Woerd
et al.~(1987). Soft X-ray modulation with a period of 14.06~s and mean amplitude of 15\%
was found towards the end of the plateau (V = 9.4) of the November 1983 superoutburst, and
oscillations with periods of 14.2 to 14.4~s were found near the peak (V = 8.8) of the October
1984 superoutburst. Extensive observations by EXOSAT of these and other VW Hyi outbursts showed no 
other instances of DNOs.

On the other hand, DNOs in the X-ray and EUV regions have been extensively observed during
SS Cyg outbursts. We will therefore briefly describe the SS Cyg observations and suggest a model 
to account for them which, mutatis mutandem, may also apply to VW Hyi.

\subsubsection{DNOs in SS Cyg}

Optical DNOs in the period range 7.3 -- 10.9~s are observed in SS Cyg during outburst (see references
in Warner 1995a), and soft X-ray DNOs are observed in the same range (Jones \& Watson 1992 and references
therein). Only recently have the two spectral regions been observed simultaneously, establishing 
the identity of periods (C.~Mauche, private communication). DNOs in the EUV region (100 -- 200 {\AA}) have
been observed in several outbursts of SS Cyg (Mauche 1996a,b, 1997, 1998) with the aid of the EUVE satellite. 
An advantage of working in the EUV is that the flux directly monitors $\dot{M}(1)$. The long expected
$\dot{M}(1) - P$ correlation has been confirmed (the ${\rm V} - P$ relationship is multivalued because the 
${\rm V} - \dot{M}(1)$ relationship is different on the rising and falling branches of a dwarf nova outburst).
The most dramatic observation, however, is the discovery of frequency doubling during the outburst
of October 1996 (Mauche 1998), producing DNOs with $P \sim 3$~s. A similar period was observed
in X-ray DNOs during the subsequent outburst (van Teeseling 1997). At the maxima of two earlier outbursts
(August 1993 and June 1994) the otherwise purely sinusoidal EUV DNOs were found to contain substantial
components of first harmonic (Mauche 1997), suggestive of incipient transition to a frequency doubled
state.

\subsubsection{The Geometry of Accretion by a Weak Field}

As a possible explanation of this behaviour we consider the nature of accretion from the inner
edge of a truncated disc into the magnetosphere of a primary with a weak magnetic field. For weak
fields and large $\dot{M}$, cooling of an accretion flow is effected by bremsstrahlung radiation. The
height $h_s$ of the shock above the surface of the primary is (Lamb \& Masters 1979; 
see equation 6.15 of Warner 1995a)

\begin{equation}
 h_s = 9.6 \times 10^7 \, M_1(1) \, R_9^{-1}(1) \, N_{16}(e) \,\,\,\, {\rm cm}
\label{height}
\end{equation}

where $N_{16}(e)$ is the electron density (in units of $10^{16}$ cm$^{-3}$) in 
the post-shock region. Using a post-shock velocity $v^2 = G M(1) / 8 R(1)$, the
continuity equation $\rho v / B \propto \rho v r^3$ = constant for fluid flow in 
dipole geometry, and a fractional area of the white dwarf surface covered by the accretion
zone $f = 0.25 (R(1)/r_0) (\delta / r_0)$, where $\delta$ is the 
width, within the DZ, of the accreting zone of the disc (Wickramasinghe, Wu \& Ferrario
1991), we find

\begin{eqnarray}
\nonumber N(e) & = & 1.48 \times 10^{14}  ({{r_0}\over{\delta}})  {\omega}_s^{2\over{3}} \dot{M}_{17}
 R_9^{-{5\over{2}}}(1) M_1^{-{1\over{6}}}(1) \\
 & & \,\,\,\,\,\, P_1^{2\over{3}}(1) (1 + {{h_s}\over{R(1)}})^{-3} \,\,\,\, {\rm cm}^{-3}
\label{ne}
\end{eqnarray}

where $P_1(1)$ is the spin period of the primary (or equatorial belt) in units of
10~s, and the radius of the inner edge of the disc (from the discussion in Section 2.2.1) is taken
to be 

\begin{equation}
r_0 = 6.96 \times 10^8 {\omega}_s^{2\over{3}} M_1^{1\over{3}}(1) P_1^{2\over{3}}(1) \,\,\,\, {\rm cm}.
\label{radi}
\end{equation}

The stand-off shock height will be within the field-dominated accretion flow if

\begin{equation}
{{h_s}\over{R(1)}} < {{r_0}\over{R(1)}} - 1.
\end{equation}

If, from equations~\ref{height}, \ref{ne} and \ref{radi}, it emerges that 
$h_s > r_0 - R(1)$ then there is no shock, and flow from the inner edge of the disc
to the primary is pressure supported and subsonic.

Three regimes can be identified:

\begin{enumerate}
\item{$h_s > r_0 - R(1)$. This will happen at low $B(1)$ and low $\dot{M}$ (see Figure 1 of
Wickramasinghe, Wu \& Ferrario 1991) and produces no shock-heated accretion zones (other than the
DZ).}
\item{$h_s < r_0 - R(1)$, but $h_s \ga 1.2 R(1)$. Here the shock-heated parts of the accretion
curtains stand well above the surface of the primary and (depending on orbital inclination) both may be visible
to an outside observer -- but only one is visible to (and hence irradiates) each surface of the disc.}
\item{$h_s \ll R(1)$. This requires relatively high $\dot{M}$, as possibly achieved at the maxima of dwarf novae
and in nova-like variables. Because of the large cross sectional area of the accretion flow, it may be of
low optical thickness along the flow, but optically thick in the transverse direction.}
\end{enumerate}

Regimes (ii) and (iii) correspond roughly to the IP structures described by Hellier (1996), Allan, Hellier \& Beardmore (1998)
and more completely by Norton et al.~(1999). These authors show that two-pole disc-fed accretion will 
lead to a single X-ray pulse per rotation if the transverse optical depth of the accretion flow is low, but
along the flow the optical thickness is high (which results in a superposition of the radiation maxima from the 
two accretion zones), whereas two pulses per cycle arise if the roles of the transverse and longitudinal 
optical depths are reversed. Among the IPs, the smaller values of $P(1)$ ($\sim$100~s) are double pulsed and
the longer period ($\sim$1000~s) systems are single pulsed. In the dwarf novae, the combination of (variable)
field strengths and $\dot{M}$ variations may be leading to successive passage through the various regimes.

Even without changes in optical depths, regime (iii) introduces geometric complications for primaries with low
fields. Equation~\ref{radi} shows that the inner edge of the disc is close to the surface of the primary.
As a result, obscuration of part or all of one hemisphere of the primary, or even of the accretion zones in
equatorial regions, can occur. For example, for $r_0 \la 1.2 R(1)$, for orbital inclinations that are at least moderate,
the primary obscures the accretion zones when they are on its far side, and the disc may obscure them for much of the phases
when they are on the near side. This leaves the upper accretion zone visible only on each side of the primary, i.e.~twice
per rotation, producing a first harmonic dominated light curve.

The observations of DNOs in SS Cyg enable this model to be made partially quantitative. Models of SS Cyg outbursts
(Cannizzo 1993) give $\dot{M}(1) \simeq 6 \times 10^{17}$ g s$^{-1}$ at maximum luminosity, but the recent HST 
parallax determination (Harrison et al.~1999) gives a distance about twice that adopted by Cannizzo, so we increase
the latter's $\dot{M}$ value by a factor of four. The EUVE observations show that
DNOs first appear, with $P \simeq 9.5$~s, at about 25\% of the maximum EUV luminosity, and that frequency doubling occurs
(from $P \simeq 5.8$~s to $P \simeq 2.9$~s) when $L \simeq 90$\% of maximum luminosity.  If we identify the turn-on  of
DNOs as the transition from regime (i) to (ii), then this occurs at $\dot{M}_{17} \simeq 6.0$. SS Cyg has an unusually large
$M(1)$ (Robinson, Zhang \& Stover 1986) -- as indeed it must have in order to allow DNOs as short as 6~s. We take
$M(1) = 1.20$, in which case $R_9(1) = 0.385$.

If we treat ($\delta/r_0$) and $\omega_s$ as unknowns, then the condition $h_s = r_0 - R(1)$, equations~\ref{height},
\ref{ne}, \ref{radi}, and the parameters for SS Cyg, produce $r_0/R(1) = 1.69$, $\omega_s = 0.87$ and $\delta/r_0 = 0.13$,
where we have made use of the additional relationship $\delta/r_0 \simeq (1 - \omega_s)$ (Wickramasinghe, Wu \&
Ferrario 1991). Too much should not be deduced from these figures -- but the fact that a physically 
sensible solution is found at all is encouraging. 
Taking a step further, keeping $\delta/r_0$ and $\omega_s$ fixed and calculating
$h_s$ when $\dot{M}_{17} = 20$ (i.e.~when the first harmonic appears), gives $r_0/R(1) \simeq 1.25$ and $h_s/R(1) \simeq 0.02$
which should certainly lead to strong geometric effects and the production of first harmonic effects
in the X-ray/EUV region.

We note that the EUV fluxes of SS Cyg and U Gem, at the ends of normal outbursts, also show dips
(Mauche, Mattei \& Bateson 2001). It will be interesting to see whether DNOs, if present at these times,
also have more rapid period increases, indicative of propellering.

\subsubsection{DNOs in AH Her and VW Hyi}

In AH Her DNOs are present on the rising branch of outburst, disappear about 0.5 d before optical maximum
and reappear about 2 d after maximum. These phases are probably more symmetrically placed about EUV or $\dot{M}$
maximum. The observations and analyses (Stiening, Hildebrand \& Spillar 1979; Hildebrand et al.~1980) are sufficient
to show that frequency doubling does not occur, which might otherwise have accounted for the disappearance.
For VW Hyi, however, we have occasional glimpses of the evolution of the DNO period down to a minimum of 14.06 s,
also with no indication of frequency doubling. The low success rate of detecting DNOs in VW Hyi for $P \la 20$ s suggests
that more complete coverage of AH Her could also be successful, and that the observed minimum at $P = 24.0$ s 
does not represent the period at which the magnetosphere is crushed to the surface of the primary.

The appearance of DNOs at 14.06 $\pm$ 0.01 s in two separate outbursts in VW Hyi (Section 4.2.1 of Paper I), 
and the fact that this is the minimum
observed period and that it occurs several days after maximum $\dot{M}(1)$, may mean that magnetically 
channelled accretion is absent at the highest $\dot{M}(1)$, but a brief and slight reduction allows the
establishment of channelled flow. In this case, with the Nauenberg (1972) mass--radius relationship
for white dwarfs, and the assumption that 14.06 s is the Kepler period at the surface of the white
dwarf, we have $M_1(1)$ = 0.702 for VW Hyi. For comparison, taking $P = 5.7$ s as the minimum for SS Cyg yields
$M_1(1)$ = 1.18.

\subsection{OY Car}

Optical DNOs in the SU UMa star OY Car, in the period range of 19.4--28.0 s, were observed
by Schoembs (1986) at the end of a superoutburst as the star fell from 12.9 to 14.4 magnitude
(at maximum OY Car has V(supermax) $\simeq 11.5$, and at quiescence V $\sim 15.5$).
Marsh \& Horne (1998; hereafter MH) observed 18 s DNOs in OY Car using the HST at the very
end of a superoutburst, just before it reached quiescent magnitude and therefore at a similar 
stage to the prominent DNOs seen in VW Hyi. Important aspects of these DNOs are
(a) the presence of two periods with amplitudes 8--20 percent, at 17.94 s and 18.16 s, the 
former being sinusoidal, the latter being weaker but having a first harmonic with 
amplitude larger than that of its fundamental; (b) flux distributions of the oscillating 
components that are very much hotter than the average spectrum, and the first harmonic
of the 18.16 s modulation is hotter than its fundamental; (c) eclipse of the oscillations
that is total and occurs during the orbital phase range $\pm$0.03, rather than the 
range $\pm$0.075 seen in the optical DNOs (Schoembs 1986), showing that the 1100--2500 
{\AA} oscillation flux is concentrated near the centre of the disc.

The duration of the eclipse of the HST oscillations at 17.94 s and the 9.07s harmonic in
OY Car is the same as that of the primary as measured at optical wavelengths (Wood et al.~1989)
and of the X-ray eclipse in quiescence (Pratt et al.~1999). This demonstrates that the 
oscillatory components are all produced very close to the surface of the primary.

Our interpretation of these observations is similar to that for VW Hyi. The 17.94 s modulation
is the rotation period of the equatorial belt of VW Hyi; the 18.16 s period is the result
of the 17.94 s rotating beam being processed by a thickening of the inner disc rotating 
progradely with a period of 1480 s. The short eclipse duration of the processed component
specifically requires the processing site to be close to the primary and not at the outer
edge of the accretion disc. The first harmonic modulation, at 9.07 s period, is either the
result of two accretion zones being visible to the thickened annulus of the disc, but
not directly to us, or is due to the irregular shape of the processing blob.

A prediction of this model, in analogy with VW Hyi, is that QPOs with a period $\sim 1500$ s
may occur in the light curve of OY Car at the end of outburst. The HST observation made by
MH had a duration of only $\sim 2250$ s and therefore (in the presence of eclipses and
other orbital modulation) was unsuitable for detection of such a QPO. 
However, the XMM-Newton X-ray observations of OY Car obtained by Ramsay et al.~(2001) about 
two days after a normal outburst showed a $\sim 2240$ s modulation characteristic of
obscuration of a central source by material circulating periodically above the plane
of the accretion disc, having large amplitude in the 0.1--1.0 keV range but no detectable
modulation in 1.0--10.0 keV. This may be a longer period evolution from an earlier
$\sim 1500$ s modulation.

Occasional dips in the optical light curve of OY Car have been observed late in outburst
(Schoembs 1986) and in quiescence (Schoembs, Dreier \& Barwig 1987) that probably
arise from obscuring matter, but no periodicities have been determined with any 
certainty. However, as Schoembs et al.~pointed out, the dips in OY Car at minimum
resemble those in WZ Sge -- for which we have a new interpretation based on our
model for VW Hyi.

It is possible that the FeII curtains, seen in HST spectra of OY Car (Cheng et al.~2000)
and in V2051 Oph, Z Cha and WZ Sge (Catal\'an et al.~1998), all of which possess DNOs, are the
result of IP-like accretion curtains rotating with periods of tens of seconds.

\subsection{WZ Sge}

\begin{figure*}
\centerline{\hbox{\psfig{figure=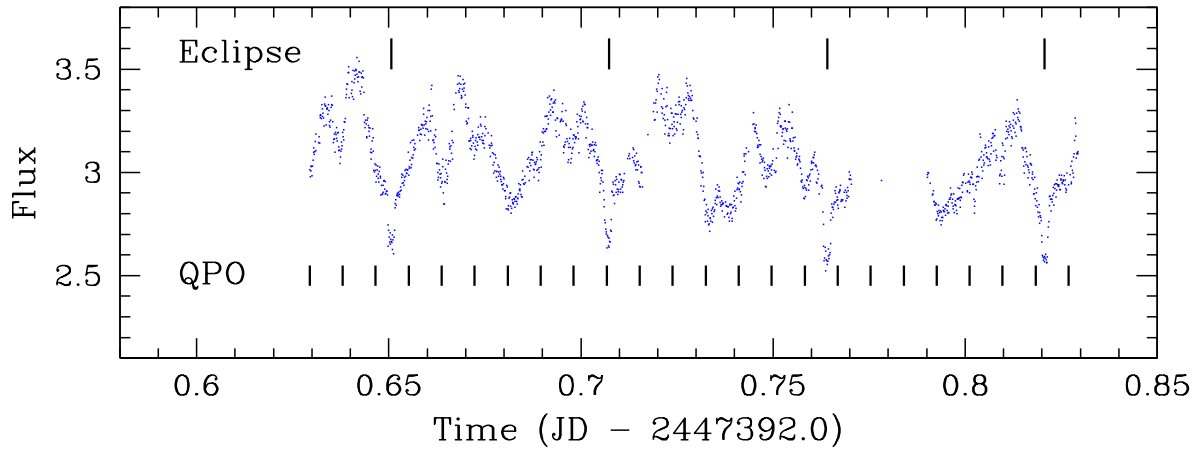,width=16cm}}}
  \caption{The light curve of WZ Sge, taken on 19 August 1988 by Dr.~J.H.~Wood. 
The eclipses are indicated by the upper vertical bars, and the QPO minima derived from the
Fourier transform are marked by the lower bars.}
 \label{wzsge}
\end{figure*}

WZ Sge is an SU UMa type dwarf nova with superoutbursts occuring two or three decades apart. Brightness
modulations at 28.952 s and 27.868 s in quiescence were discovered by Robinson, Nather \& Patterson (1978),
who found that in general either one or the other, but occasionally both were present.
Many studies have been made of these oscillations, reviewed and extended in Patterson et 
al.~(1998), with frequent claims that the 27.87 s oscillation and some associated sidebands
are evidence for a rotating magnetic accretor (Patterson 1980; Warner, Tout \& Livio 1996;
Lasota, Kuulkers \& Charles 1999). Important support for this model has come from the 
discovery of X-ray modulation at 27.87 s (Patterson et al.~1998), but analysis of HST 
observations (which also are dominated by the 27.87 s modulation) has
shown that a simple magnetic accretor does not account for all of the observed
periodicities (Skidmore et al.~1999), and that non-radial pulsation of the white dwarf
primary could be a viable mechanism. Nevertheless, the observed rotation velocity
of the primary in WZ Sge, $v \sin i = 1200$ km s$^{-1}$ (Cheng et al.~1997b), together
with $i = 70^{\circ}$ and $M_1(1) = 1.0 \pm 0.2$ (Cheng et al.~1997b; Spruit \& Rutten
1998), gives $P(1) = 28 \pm 8$ s, which is supportive of the model of rotational
modulation from a magnetic accretor.

One of the problems with all models of the WZ Sge oscillations is the 28.95 s period, which
does not appear related in any obvious way to the 27.87 s period and its optical sidebands. 
But by analogy with VW Hyi and OY Car we now have a simple explanation -- which was first
suggested in essence by Patterson (1980), but now we have a more clear idea of what a 
blob rotating in the disc consists of. Patterson's idea was that the 28.95 s signal arises
from a 27.87 s rotating beam processed by a blob revolving in the accretion disc at the beat
period of $733 \pm 15$ s. Lasota et al.~(1999) claimed that to have such a period the reprocessing site
would have to be far out in the disc and suggested the outer rim as a region that might be
capable of exciting and sustaining QPOs. 

We prefer, however, to adopt the VW Hyi model, with a 
QPO travelling wave in the {\it {inner disc}}. WZ Sge is unique, however, because it then
becomes the only dwarf nova showing persistent DNOs and QPOs in quiescence. This implies a
primary with a magnetic moment intermediate between those of normal dwarf novae and those of
IPs, and also an accretion history that spun the primary up to its present rapid rotation
rate. Again by analogy with VW Hyi it should be kept in mind that there may be an 
equatorial belt generating or enhancing the magnetic field and not firmly attached to the 
primary. That this is even probable is seen from the value of $P/\dot{P} \sim 1 \times 10^5$ y
found for the 27.87 s oscillation in the 1976--1978 data by Patterson (1980), which is too
short a time scale for it to involve the whole of the primary. 
Again in analogy with VW Hyi, no first harmonic of the 27.87 s DNO is seen, but a harmonic
of the reprocessed signal at 28.96 s is observed at 14.48 s (Provencal \& Nather 1997).

Our QPO processing model for WZ Sge requires there to be a thickened region of the disc
rotating progradely with a period of 744 s, which is the beat period of 28.952 s and 27.868 s. 
As well as processing radiation from the central 
source, this feature, when of sufficient amplitude, might be expected to obscure part of the
central region in this high inclination system. We will demonstrate that this is in fact
the case.

From the first light curves obtained for WZ Sge (Krzeminski \& Kraft 1964) observers have
commented on the presence of a dip of unknown origin in the light curve in the vicinity 
of orbital phase 0.3 (see, e.g., Fig 3.3 of Warner 1995a). The base level of this dip is 
often as deep, but rarely much deeper, than the minimum in the WZ Sge light curve which
gives it its unusual ``double hump'' or W UMa-like profile. The dip at phase $\sim 0.3$
is so prominent that, although other, lesser dips are present, the latter have not been 
taken notice of. However, armed with the expectation of repetitive dips spread 744 s apart, 
it is immediately obvious (a) that many of the published light curves have such dips
(e.g.~the upper light curve in Fig 3.3 of Warner 1995a) and (b) the reason that they are 
not conspicuous is because they do not, or cannot, occur near phase 0.5 where
the light is already reduced by another cause.

\begin{figure}
\centerline{\hbox{\psfig{figure=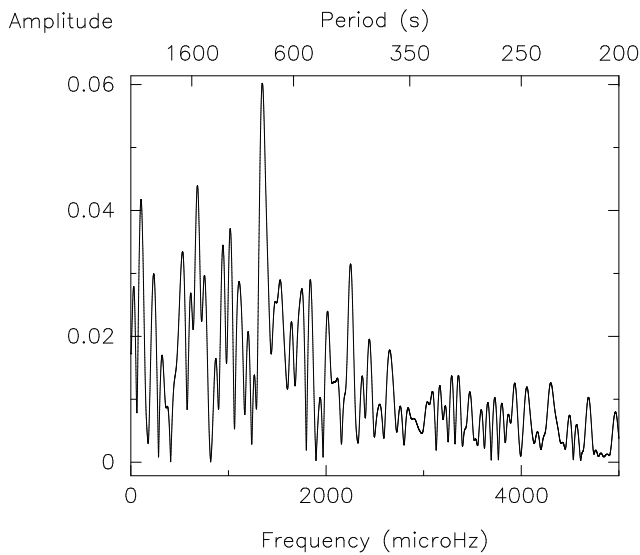,width=8.8cm}}}
  \caption{The Fourier transform of the light curve shown in Figure~\ref{wzsge} after prewhitening.
The highest peak is at 742 s.}
 \label{wzsge_tranf}
\end{figure}

We reproduce in Fig.~\ref{wzsge}~a light curve of WZ Sge obtained by Dr.~J.H.~Wood through a B filter
on 19 August 1988 using the Stiening multichannel photometer on the 82-in reflector
at the McDonald Observatory. The original light curve, at 1 s time resolution, has been
binned to 10 s. In addition to the eclipses every 81.6 min there are minor dips
appearing at all phases other than around 0.5. Even by eye there is a suggestion of 
periodicity. This is confirmed by the Fourier transform of the light curve, shown in Fig.~\ref{wzsge_tranf},
in which we have prewhitened the light curve at the two strongest periodic components (the first
and third harmonics of the orbital period) and which shows a prominent peak at $742 \pm 1$ s.
No DNOs are detectable in this light curve, showing that the QPO exists 
independently of the presence of DNOs.

Our impression, inevitably subjective, is that Fig.~\ref{wzsge} shows only recurring
dips and not the peaks seen in VW Hyi. This is in accord with the expectation (Section 3.2.1) that
reprocessing from the wall will be largely hidden by the primary in a high inclination system.
In addition, the primary in WZ Sge is cool ($\sim 15\,000$ K) and the accretion luminosity onto
it is low, so relatively little radiation is available for reprocessing.

Much of the `noise' on each side of the 742 s signal in Fig.~\ref{wzsge_tranf} is sidebands
generated by the severe amplitude modulation of the recurrent dip in Fig.~\ref{wzsge}. The
signal itself is the average amplitude and is much smaller than the extreme depths of the
dips. In our new interpretation of Fig.~\ref{wzsge}, all of the major features 
arise from eclipse, the broad dip at orbital phase 0.5, and QPO dips.
There is very little short time scale flickering.

It is usually thought that, with the (possible) exception of WZ Sge, no
dwarf novae have DNOs in quiescence.  We draw attention to the
Einstein hard X-ray observation of HT Cas in quiesence on 23 Jan 1980,
which showed a significant period of 21.88 s and an equal amplitude first
harmonic (Cordova \& Mason 1984).  This is similar to the optical DNO
periods seen during outburst.  Warner (1995a) has pointed out that
spectroscopic observations of HT Cas in quiescence are suggestive of
stream overflow and impact onto a magnetosphere.

\subsection{UX UMa}

Optical DNOs with periods near 29 s were discovered in the nova-like
UX UMa by Warner \& Nather (1972) and more extensively observed and
analysed by Nather \& Robinson (1974; hereafter NR).  More recently  UV
DNOs have been discovered  by Knigge et al.~(1998a) with HST observations.
A number of points relevant to the content of this paper can be extracted
from the observations of UX UMa:

\begin{itemize}
\item{In analysing the flux distribution and eclipse observations obtained
by HST, Knigge et al.~(1998b) found that they could not get a good fit with
standard theoretical discs.  One option they point out, which improves the
fit, is to truncate the disc by removing the inner regions out to a few
white dwarf radii.  This also improves the fits in the nova-like IX Vel
(Long et al.~1994), which is another system showing $\sim$26 s DNOs (Warner,
O'Donoghue \& Allen 1985).  Such results are certainly compatible with the
model of magnetically controlled accretion in the inner disc.}

\item{Knigge et al.~(1998a) found a large shift in DNO oscillation phase
during eclipse, similar to that seen by NR.  In fact, in
the two HST eclipse observations the phase shift was more than 360$^{\circ}$, 
suggesting that the mean period derived for the DNOs before
eclipse had changed at some time through eclipse. There are steps in the O
- C curve through and after eclipse that resemble the discontinuous jumps
in period seen in TY PsA and VW Hyi.}

\item{In the O - C curves of NR and the second eclipse of Knigge et al.~(1998a) 
there are oscillations - with a quasi period $\sim$650 s in the former
and $\sim$400 s in the latter (at the end of and after eclipse).  These are on
a time scale suggestive of modulation by reprocessing from a travelling
blob.}

\item{DNOs in UX UMa are not always present.  Their absence could  mean
that although present they are hidden from view from our perspective, or it
could be that an increase in $\dot{M}$ has crushed the magnetosphere down to
the surface of the white dwarf, or it could be the result of a slight
decrease in $\dot{M}$, which results in an increase in $r_0$ and cessation
of accretion onto the primary because of propellering or at least
prevention of steady accretion. In principle, distinction between the last
two could be obtained if there is found to be a correlation between EUV
flux and presence of DNOs.  The damming up of accretion in the inner disc
of an IP leads to an instability (Spruit \& Taam 1993) which can produce
short-lived outbursts (Warner 1995c) of the kind seen in the nova-like RW
Tri (Still, Dhillon \& Jones 1995), and which may be used to infer
magnetically-truncated discs. Again, it would be interesting to see if
there is a minor outburst in the EUV at the end of a period of DNO
absence. }
\end{itemize}

\subsection{V2051 Oph}

Steeghs et al. (2001: hereafter S2001) have observed both continuum and spectral
line modulations in V2051 Oph towards the end of a normal outburst (at V $\simeq$ 14.8).
Continuum oscillations at 56.12 s and its first harmonic at 28.06 s are interpreted as
originating from the white dwarf primary. The length of the observation (35 min) was insufficient
to determine whether this modulation has the typical period variation of a DNO, but an earlier
observation of a 42 s period (Warner \& O'Donoghue 1987) at a brighter phase (V $\simeq$ 13.2)
suggests that it does. A period of 29.77 s was also observed, which has a beat period of 488 s
with the 28.06 s modulation. The absence of a sideband at 26.53 s shows that, as in our observation
of VW Hyi (Figures 18 and 19 of Paper I), the longer period is a reprocessed signal from a blob moving with the period
of 488 s. Evidently there are two accretion regions in the equatorial belt of V2051 Oph, which itself
has a rotation period of 56.12 s.

Periods of 28.06 and 29.77 s (but not twice their values) were also observed in the Balmer lines.
Their similar amplitudes result in an apparent almost 100 percent amplitude modulation at 488 s. There
is a weak indication of the 488 s period (which we recognise as a QPO) in the continuum light curve which, 
if correctly identified, implies that the maximum of the Balmer line oscillations (i.e., the time
when the 28.06 s and 29.77 s modulations are in phase) corresponds to minimum of the QPO.

V2051 Oph is a high inclination system ($i \simeq 83.5^{\circ}$: Baptista et al.~1998), which leads
to strong back--front asymmetry in the system, obscuration of the inner parts of the accretion disc
by the primary, and the possibility of partial obscuration of the primary by a travelling wave in the 
inner disc. We would expect that the dominant continuum signal at $P(1)$ would result from reprocessing
from the far side of the disc. Reprocessing from the travelling blob would also be seen at maximum
advantage when the blob is on the far side except that it is partially obscured by the primary if the blob 
is in the inner region of the disc: taking $P(1) = 56$ s and $M_1(1) = 0.78$ (Baptista et al.~1998) gives
$r_0 \simeq 3.0 R(1)$ from equation~\ref{radi}, so by analogy with VW Hyi vertical thickening of the 
disc may be excited just beyond $r \sim 3 R(1)$.

S2001 find a strong wavelength dependent phase shift across the emission line profile, in the sense 
that oscillation maximum moves from blue to red across the profile every 29.77 s. At the 28.06 s period
no strong systematic effect is observed. There is similarity with what is seen in DQ Her, where the
phase dependence has the opposite sign, i.e., the wavelength of maximum amplitude moves from red
to blue across the profile (Chanan, Nelson \& Margon 1978; Martell et al.~1995). In DQ Her the inclination
is so great that only the far side of the disc is visible over the front edge of the disc, so a 
prograde rotating beam, as seen by us, can only illuminate material that is sequentially receding,
moving transversly, and approaching, with little or no part of the return path visible. The earlier of
these observations of DQ Her was very helpful in confirming the model of a rotating reprocessing beam.

The different behaviour in V2051 Oph can be interpreted as follows. The two rotating beams produce
a 28.06 s signal from ionization and recombination of the axially symmetric part of the optically thin
upper atmospheric regions of the disc in its inner parts (i.e., just beyond $r_0$). Because
of obscuration by the primary, and optical thickness effects in the material at radial velocity
extrema (see below) there is not a clear or complete S-wave produced by spectra folded at this
period (see Fig.~7 of S2001) -- in particular, the passage of the S-wave from red to blue is reduced
in intensity because of obscuration by the primary (e.g., at $r_0 = 3 R(1)$, the Keplerian rotation
is $v_0 \simeq 2200$ km s$^{-1}$ and obscuration by the full diameter of the white dwarf would
remove the return section of the S-wave for velocities in the range $+700 > v > -700$ km s$^{-1}$). 
It should also be kept in mind that the illuminating beam may be very broad and therefore the profile 
of the reprocessed line emission will be broadened, asymmetric and have a velocity amplitude
smaller than $v_0$.

Reprocessing from the travelling blob is similarly affected: gas streams through the region
of vertical thickening at the Keplerian velocity, the elevated highest regions will generate
an emission line spectrum, and illumination by the beam generates a modulation at the `synodic'
period of 29.77 s. Because the region is optically thin in the visible, but the vertical
density falls roughly exponentially, the largest emission measure will occur at a height where
the gas first becomes optically thin. When the travelling wall is at inferior conjunction (i.e.,
`back-lit' by the rotating beam) the axially symmetric component and the component from the blob
are in phase, agreeing with the appearance of largest amplitude of the Balmer emission line modulation
at QPO minimum. When the wall is at quadrature the larger physical path length in our direction
will push the height of optical thinness higher, where the gas density and emission measure are lower,
reducing the emission line intensity. This and the velocity-averaging effect of a broad illuminating beam
and an extended blob, reduce the apparent velocity amplitude of the S-wave from the expected
$\pm 2200$ km s$^{-1}$. In Fig.~8 of S2001 the emission appears to extend to $\pm 1800$ km s$^{-1}$.
We feel, therefore, that a travelling wave at $\sim 3 R(1)$ in the disc, rather than a 
blob moving at Keplerian velocity at $\sim 12 R(1)$, as advocated by S2001, is compatible with 
the observations. It will certainly be easier to account for the readily visible $\sim 29$ s 
oscillations in the Balmer line flux if the reprocessing site is at $3 R(1)$ than if it is 
at $12 R(1)$.

Finally, we note the unusual light curve of run S3320 shown in Fig.~1 of Warner \& O'Donoghue
(1987). The dips in the first two thirds of this light curve, which was of V2051 Oph during 
outburst, resemble those in WZ Sge (Fig.~\ref{wzsge}). We suggest that there was a large
amplitude QPO present, with a period near to one third of the orbital period (i.e., $\sim 1800$ s) 
which decayed steadily over the $\sim 7.5$ h duration of the run. We have confirmed that there
were no DNOs in this run, although the 42 s DNO was present on the previous night. This is a further
example of the independence of DNOs and QPOs.

\subsection{V436 Cen}

DNOs near 20 s period were found in V436 Cen during a superoutburst (Warner 1975) and
analysed by Warner \& Brickhill (1978). The period (obtained by periodogram analysis 
of successive sections of the light curve) varied cyclically over a range $\sim$19.3--20.3 s
This effect has not been previously understood, but noting that there are QPOs present
in the light curve (see Fig.~1 of Warner 1975) which have periods in the range
400--500 s, we suggest that a `direct' DNO with period $\sim$19.3 s is sometimes dominant, 
and at other times its `reprocessed' component at periods 20.1--20.3 s dominates. This will be analysed
more completely elsewhere.

\section{Concluding Remarks}

The principal observational and interpretative points made in our two papers are:

\begin{itemize}
\item{Optical DNOs over the range 14 -- 40 s have been observed in VW Hyi outbursts 
-- previously the shortest periods had been detected only in X-Ray observations.}
\item{Over a time of about 12 hours at the end of outburst the DNOs in VW Hyi 
increase rapidly in period and then recover to shorter periods.  }
\item{QPOs are often present during outbursts.  For the first time we have detected a 
significant evolution of QPO period as the luminosity at the end of outburst 
decreases.}
\item{We develop the Low Inertia Magnetic Accretor model, in which accretion is 
taking place onto a weakly magnetic white dwarf, spinning up an equatorial band 
which thereby enhances the intrinsic magnetic field. In this way, magnetically 
channelled accretion can occur even on low field CV primaries.}
\item{Frequent DNO period discontinuities are interpreted as magnetic reconnection 
events.  The observed reduction of coherence in DNOs late in outburst is a 
consequence of the more rapidly changing rate of mass transfer through the inner 
disc, which causes the inner radius of the disc to move outwards, preventing an 
equilibrium (as in an intermediate polar) from being attained.}
\item{Because the rate at which the inner disc radius moves outwards is greater than the 
equatorial belt can adjust to, there is an inevitable stage of ``propellering'' in which 
much of the accreting gas is centrifuged away from the primary.  This corresponds to the 
observed phase of rapid deceleration of the DNOs.  At this time, the rate of mass 
transfer onto the primary, as indicated by the EUV flux, falls almost to zero.}
\item{The frequency doubling of DNOs observed in the EUV of SS Cyg is attributed to 
optical thickness and geometric effects that arise from the LIMA model.}
\item{The QPO modulation is interpreted as a prograde travelling wave (or ``wall'') near 
the inner edge of the disc. The luminosity variations come from obscuring and 
reprocessing of radiation from the central region of the disc (and primary), rather 
than as variations intrinsic to the disc oscillation itself. A possible excitation 
mechanism, associated with the magnetic reconnection mechanism, is proposed.}
\item{The observed modulation of the DNOs at the QPO period -- leading at times to a 
`QPO sideband' -- arises from interception of the rotating DNO beam by the 
travelling wall.}
\item{Applied to other CVs, the models account for the pair of DNO frequencies 
observed in OY Car during outburst (MH),  the persistent pair of DNOs seen in 
WZ Sge during quiescence (in which we demonstrate for the first time the 
existence of a QPO frequency which is the difference frequency of the DNOs), the 
pair of DNOs seen in V2051 during outburst (S2001), and the puzzling behaviour 
of DNOs seen in V436 Cen during outburst.}
\end{itemize}

We have proposed that the QPOs are due to the interception of radiation by a travelling 
wall. In this connection the QPOs in SW UMa observed during a superoutburst  (Kato, 
Hirata \& Minishege 1992) are very informative.  Near the minimum of each 
sinusoidal $\sim$370 s modulation, where in our model the wall would be partially obscuring  
the inner disc, a dip of depth $\sim$0.1 mag lasting for $\sim$70 s was seen.  The duration is 
appropriate for partial eclipse of the primary by the highest part of the travelling wall. 
Kato et al.~(1992) suggest obscuration by a body of horizontal dimension $> 6 \times 10^9$ cm rotating at 
the Keplerian period, which would place it at $\sim$7.5 $R(1)$.  The orbital inclination of SW 
UMa is not known, but is believed to be $\sim 45^\circ$ (Warner 1995a).  To obscure the primary 
would therefore require a vertical thickening  $\ga 7 \times 10^9$ cm.
We suggest that a QPO wall at a radial distance of the disc $\sim 1.5 R(1)$ 
and vertical height $\sim 1 \times 10^9$ cm would be a more realistic proposition.

A deeply eclipsing CV with large amplitude QPOs is required to help locate the radial 
distance of the travelling wall. Unfortunately WZ Sge does not satisfy this requirement 
(Fig.~\ref{wzsge}) because there are other modulations that affect the shape of eclipse and make 
interpretation difficult.

The possible success of the LIMA model developed here is seen as evidence for CV 
primaries with magnetic fields weaker than those of IPs -- an extension to $B \la 1 \times 10^5$ G, 
where the field is not strong enough to lock the envelope to the interior, but is powerful 
enough to cause some channelled accretion flow. Isolated white dwarfs exist with $B < 
10^5$ G (Schmidt \& Smith 1994) so we can expect CV primaries also to be represented 
at this level. Such weak fields are currently not measurable directly, but the DNO 
phenomena provide indirect evidence for their existence.

That some parameter, taking different values in different systems, lives in CVs is seen 
from the fact that otherwise similar CVs can have strong or absent DNOs. In our 
interpretation this parameter is $B(1)$, and the latter stars have small or zero values of it.  
Variations of $B(1)$, and $\dot{M}$, and differences of $M(1)$, combine to produce the range of 
DNO phenomena.  The apparent absence of DNOs in any IP is in accord with the LIMA 
model -- fields strong enough to generate the canonical IP behaviour should prevent 
slippage of the equatorial belt. Similarly, we would not expect a hot equatorial belt
to be observed after an outburst of an IP.

Not all high $\dot{M}$ discs show QPOs, again showing the existence
of a hidden parameter -- which might again be $B(1)$ and the ability to excite
waves in the disc.

It has not escaped our notice that the DNO and QPO phenomena described here have 
apparent analogues among the accreting neutron stars.  For example, the behaviour of the 
O--C diagram for the $\sim$5 Hz QPOs in the Rapid Burster (Dotani et al.~1990) is similar to 
that in CVs. The kHz DNOs and their companions at not quite constant frequency 
difference (e.g. van der Klis 2000) resemble the double DNOs in VW Hyi, OY Car and 
WZ Sge.  The frequencies depend on luminosity, as in CVs. Furthermore, the VW Hyi 
QPO/DNO period ratio of $\sim$15 (Figure 3 of Paper I) is the same as that seen in X-ray 
binaries -- indeed, our VW Hyi data fit on an extrapolation of Figure 2 of Psaltis,
Belloni \& van der Klis (1999) to frequencies two orders of magnitude lower than seen
in the X-ray systems. This area, where similar 
phenomena appear, separated by the gravitationally determined ratio of time scales 
$(\bar{\rho}_{ns}/ \bar{\rho}_{wd})^{1\over{2}}$, remains to be explored.

\section*{Acknowledgments}

Part of this work was carried out while BW was visiting the Astrophysics
Groups at Keele University and the Australian National University. He expresses
his gratitude to both for their support and hospitality.
Financial support for this work came from the University of Cape Town. We are
grateful to the anonymous referee for suggestions that have led to 
improvements of presentation of this and Paper I.

\end{document}